\documentclass{eas}
\usepackage{graphicx}

\begin{document}

\title{ELSA and the frontiers of astrometry}

\author{Anthony G.A.\ Brown}
\address{Sterrewacht Leiden, Leiden University, P.O.\ Box 9513, 2300 RA Leiden,
The Netherlands}

\begin{abstract}
  ELSA stands for the ambitious goal of `European Leadership in Space
  Astrometry'. In this closing contribution I will examine how the ELSA network
  has advanced this goal. I also look ahead to the time when the Gaia data will
  be published and consider what needs to be done to maintain European
  leadership.
\end{abstract}

\maketitle

\section{Impact of ELSA}

First, what does it take to establish and maintain a leadership position in the
field of space astrometry? I believe the following ingredients are important.
The basic condition is of course to have a wide scientific community interested
in the data that can be obtained from dedicated astrometric surveys. The
interest should not just be in the classical applications of astrometry (i.e.,
catalogues of star positions) but in pushing the limits of achievable
astrometric, photometric and radial velocity accuracies in order to expand our
horizons and learn more about our universe. This strong scientific interest
should be complemented by an expert community of astronomers capable of
designing and delivering missions such as Gaia. Because of the nature of space
astrometric missions, in terms of hardware (spacecraft and payload design) and
software (data processing), these experts should welcome contributions from the
fields of engineering, mathematics, and nowadays, computer science. The third
important ingredient is the existence of competent industrial partners capable
of delivering the very complex satellites required for space astrometry.
Finally, it is crucial that there are good contacts between these three
communities. The scientific users of the space astrometric survey have to
understand the limitations of what can be achieved, while the space astrometry
experts should not work in isolation but be motivated by the science that will
be done with the data they produce. The latter will also very much facilitate
the transmission of scientific requirements to ESA and its industrial partners.

The existence of a scientific community in Europe interested in space astrometry
is aptly demonstrated by the numerous contributions to this volume, covering a
broad range of science topics, that show the eager anticipation of the Gaia
data. The long astrometry tradition in Europe combined with the efforts to
realize the Hipparcos mission have led to a world leading community of space
astrometry experts. The Hipparcos experience was also incorporated within ESA
and its industrial partners. How has ELSA contributed to maintaining and
extending the four above mentioned components of Europe's leading role in space
astrometry?

\paragraph{Maximizing the scientific return from Gaia data}

If the Gaia catalogue with its 1 billion sources --- including high accuracy
astrometry, photometry, and radial velocities for over 100 million sources ---
would land on one's desk today it is not obvious how to most efficiently exploit
this huge data set. Creative thinking and advance preparation is required in
order to be ready to apply novel and efficient data analysis methods to the Gaia
results. Moreover, the unprecedented levels of accuracy make it essential to
think about possible complications in the interpretation of the data. Eight of
the ELSA research projects were directly relevant to these areas. The work
described by V\'aradi makes a strong case for publishing per-CCD photometry for
the (short-period) variable stars observed by Gaia, and illustrates how the
survey requirements evolve in response to preparatory research. Re Fiorentin
provides an excellent example of how one can do current science as well as
prepare for the Gaia data by analyzing the combination of existing astrometric
(GSC-II), photometric (SDSS), and spectroscopic surveys (RAVE). The works of
Santoro and Saguner illustrate how Gaia has stimulated new theoretical research
into stellar structure and atmospheres as well as preparatory ground-based
observational campaigns, which can already be exploited scientifically. Further
preparations for the scientific analysis of Gaia data are described in the
contributions by Czekai and Belcheva which focus on the non-trivial question of
how to use the Gaia data to study our galaxy or the Magellanic clouds as a
whole. These works also strengthened the Gaia universe model (see contribution
by Luri) leading to more reliable simulations of the sky for mission and
scientific preparations. The investigations by Pasquato into the effects of star
spots on the astrometry from Gaia highlight the importance of re-examining the
established methods of interpreting observational data when the accuracies are
pushed to the limit.  Finally the research into inversion methods by Oskiewicz
has stimulated much discussion about the way in which the results from Gaia
should be transmitted (more on this topic below).

\paragraph{The next generation of space astrometry experts}

One of the central aims of ELSA was of course to transfer the expertise from the
generation involved in the Hipparcos project to a new generation who will be
involved in the preparation and running of the Gaia mission. The network has
been very successful in this respect, not only by directly training a new
generation of space astrometry experts but also by serving as a focal point for
research in support of the Gaia mission preparations. This is most clearly the
case in the areas of the astrometric solution and the radiation damage
mitigation. The AGISLab package developed by Holl is an essential tool in the
development of enhancements to the astrometric iterative solution and will be
used to predict the complicated behaviour of the astrometric errors and the
correlations between them. The conjugate gradient solver for the iterative
astrometric solution was first implemented and tried in AGISLab before it was
transferred to AGIS. This led to a large improvement in the speed of the
astrometric solution as described by Bombrun which in turn facilitates research
into the properties of the solution itself. The efforts by Risquez resulted in
an accurate dynamical model of the Gaia spacecraft attitude which can be used to
furnish realistically simulated attitude data for mission preparations. During
the mission such a tool can aid the understanding of effects seen in the
attitude reconstructed from the observations (see also the contribution by Van
Leeuwen). Finally, in the case of Gaia there is no practical possibility to have
two independent consortia carry out the data processing, yet it remains
important to verify of the reliability of the Gaia catalogue through independent
checks. This task is made possible by the work described by Abbas which will
lead to an independent astrometric data reduction for a subset of the Gaia
sources.

An important problem for the Gaia mission that was identified already at the
time the mission was conceived is the issue of radiation damage to Gaia's CCDs
caused by Solar wind protons. This leads to increased charge transfer
inefficiency in Gaia's detectors and will affect all the data collected by Gaia
by causing systematic errors that have to be accounted for. Again, ELSA served
as a focus for efforts to mitigate the effects of radiation damage. The work by
Prod'homme will result in the most sophisticated model to date of the effects of
radiation damage on the charge collection and read-out process in CCDs. The
contribution by Weiler describes how this work has stimulated the development of
fast analytical models to describe radiation damage effects. Such models will
play a central role in the data processing for Gaia.

\paragraph{Strengthening the partnership with industry}

The demands of the Gaia mission are pushing the limits of what can be achieved
with conventional CCD technology and input from industrial partners has proved
essential in the studies of radiation damage mitigation. The supplier of Gaia's
CCDs, e2v, supported with their expertise the development of microscopic models
of the electron distribution within a Gaia CCD pixel (work by G.\ Seabroke at
Open University) which led to much improved modelling at the pixel level (see
contribution by Prod'homme). The modelling work and the development of the
radiation damage mitigation scheme for Gaia also relied heavily on the
laboratory experiments that EADS-Astrium carried out with irradiated Gaia CCDs
(see contribution by Pasquier). In turn the industrial partners were pushed by
the constant questioning from the scientist to enhance their own understanding
of radiation damaged CCDs. The new knowledge on CCDs was absorbed by ESA as the
facilitating partner and this will certainly be of great benefit to future
European space missions.

\paragraph{Networking and bringing in new expertise}

The opening and closing conferences of the ELSA network were aimed at bringing
together the scientist that will use the Gaia data and the community that will
produce the data. Where the opening conference served to provide the ELSA
fellows with the scientific motivation for their work, the conference of which
this volume forms the proceedings served to show the European astronomical
community how the Gaia data will be produced. I was pleased to hear an
astronomer not involved in Gaia state; ``I'm convinced I came to the right
conference!''. These important contacts between `user' and `producer' were also
fostered within the ELSA network, for example through the use of AGISLab in the
study of Gaia's capabilities with respect to variable stars.

Lastly, the ELSA network also successfully brought much needed outside expertise
into the astronomical community. Examples are the engineering expertise of
Prod'homme and the industrial partners, the mathematical expertise of Bombrun
and the knowledge of high performance computing that is brought in through the
work described by Fries in this volume.

\section{Future proofing the Gaia data}

Gaia will provide an unprecedented stereoscopic map of our Milky Way and the
nearby universe. The catalogue will contain over 1 billion stars, $\sim300\,000$
solar system objects, millions of galaxies, $\sim500\,000$ quasars and thousands
of exoplanets. For all these objects accurate astrometry, photometry, and (for a
subset) spectroscopy will be available as `basic' data. In addition the
classification, variability characterization, and astrophysical parameters of
each object will be provided.  When this catalogue is `finished' around 2020 and
combined with other large sky surveys it will become {\em the} astronomical data
resource for decades thereafter, representing a tremendous discovery potential.

However, I strongly believe that the true potential of the Gaia data can only be
unlocked if we take an ambitious and innovative approach to data publication and
access, including the provision of advanced data analysis tools. This means that
we should not plan the catalogue publication based on what we can image is
possible with current resources but rather base ourselves on what is possible
after 2020. I therefore advocate the following guidelines:

\paragraph{Publish early and publish often} The experience from the current
large sky surveys, notably SDSS and RAVE, has shown that early and regular
releases of the data are a very successful approach to survey publication. As
pointed out in the contribution by Juric, the astronomical community very much
appreciates and makes heavy use of the early releases (even if they are not as
polished as one would like); the survey producers themselves benefit from
immediate user feedback allowing them to correct important errors early on; it
enables the rapid reaction by the community to new discoveries, and will
facilitate synergies between Gaia and other projects. In this context it is
worth keeping mind the enormous synergy possible between Gaia, LSST and
Pan-STARRS, where especially in the case of LSST there is a smooth connection to
the Gaia survey in terms of the astrometric and photometric accuracies achieved.
The goal should thus be to set an ambitious publication schedule with a first
Gaia data release foreseen as soon as the sky has been surveyed once and an
all-sky catalogue of positions, and broad-band photometry can be released ($G$,
$G_\mathrm{BP}$, $G_\mathrm{RP}$). Such a trivial sounding data release (no
parallaxes, motions, or detailed astrophysical characterization yet) will in
fact constitute the highest spatial resolution all-sky map ever produced and I
am convinced that it will lead to exciting discoveries. Such a release would in
addition greatly enhance target selection for surveys that specifically aim at
complementing Gaia with, for example, high resolution spectroscopy for chemical
abundance determinations. This release should be followed with regular releases
containing ever more, more complex, and better data.

\paragraph{Keep raw data, calibration data, and processing software available}
The contribution by Van Leeuwen shows how better insights into the attitude
modelling for Hipparcos combined with present-day computing power enabled a
higher quality re-processing of the entire Hipparcos data set. This resulting
new version of the Hipparcos catalogue features very much reduced error
correlations and improved astrometric accuracies (by up to a factor of 4) for
the bright stars. This is the best illustration of the fact that the raw Gaia
data, all the calibration data, and the processing software should be stored
such that they are permanently accessible and readable, just as the catalogue
itself will be. It is in fact a prerequisite for the next guideline.

\paragraph{Facilitate (re-) processing of the (raw) data} Already in the case of
Hipparcos there are numerous examples of the re-processing of the data, notably
to improve the astrometry of binaries and very red giant stars. Other examples
include the re-processing of intermediate data for groups of stars in order to
derive a common radial velocity or parallax, the re-processing of data for
objects that are discovered or confirmed to be binaries following a data
release, or the re-determination of astrophysical parameters for stars following
improvements in atmosphere modelling. In principle also for Gaia the
re-processing of {\em all} the raw data might be warranted at some point in the
future. In addition to the re-processing of the data the Gaia archive should
also facilitate very complex operations on large chunks of the catalogue (say an
all-sky search for stellar streams). Both these aims may be best served by
implementing the idea of `bringing the processing to the data' by offering users
a virtual machine at the data centre hosting the Gaia archive. On this machine
one could code whatever analysis or processing algorithm is called for and run
it in a way specified by the user.

\paragraph{Make the archive `live'} A concept closely related to the previous
item is that of making the Gaia data archive a `living entity'. By this I mean
that it should be possible to incorporate new information into the catalogue.
Examples are complementary ground-based spectroscopy, updated classifications or
parameterizations of stars based on independent information, better distance
estimates for faint stars, etc. In addition the Gaia archive should seamlessly
integrate with other large sky surveys including ones not foreseen at the time
of the Gaia data publication. As an example, it should be possible to query the
catalogue for sources brighter and fainter than the $G=20$ survey limit of Gaia,
where behind the scenes the work is done to combine Gaia and, for example, LSST
data.

\paragraph{Don't publish a catalogue} This guideline seems in complete
contradiction to the discussion above but is meant to capture in a few words the
ideas presented in the contribution by Hogg \& Lang in this volume. I think
serious efforts should be invested in their proposal to consider a `catalogue'
as only our best current model that explains all the raw data and to publish the
data such that users can test their hypotheses (almost) against the raw image
pixels. I do not expect that this approach will be possible anytime soon but it
should not stop us from being ambitious and trying.

\section{What about ELSA-II?}

Is there a need for a follow-up to the ELSA research activities, i.e.\
investigations at the interface of astronomy, engineering, computer science and
mathematics? I think the answer is yes and that there are in fact at least two
different topics an `ELSA-II' could focus on. The Gaia data will surely
stimulate demands for future space astrometry missions that can open up new
wavelength domains to precision astrometry, push the astrometric accuracy
limits, or extend existing accuracies to much fainter stars. All these would
require technological innovations to break the current scaling laws that would
prevent us from simply building a Gaia$++$. A second and possibly more important
avenue would be to pursue the ideas outlined in the previous section. Questions
to investigate would be: how does one take care of the data curation for Gaia,
including software and calibration data, such that the stored data are easily
retrievable in the future? How do we best go about bringing the processing to
the data or combining Gaia data with other archives in a transparent manner? How
does one implement the testing of hypotheses against the raw data? Is this at
all practically possible? Note that Pfenniger argues in his contribution that
modelling the Galaxy may be best achieved through a very large particle number
$N$-body simulation, possibly containing a particle for every star in the
Galaxy. How does one decide for such models which is best? Do we convert these
models to predictions of the Gaia pixels?

Although these research topics are oriented toward the longer term future they
will generate many spin-offs that can be applied immediately to existing
astronomical instruments, astronomical data, or in industry, thus maintaining
and extending European Leadership in Space Astrometry.

\begin{acknowledgements}
  I thank William O'Mullane and David Hogg for discussions that shaped the ideas
  presented in this contribution. Special thanks go to Stefan Jordan for
  providing the nice portraits of the ELSA fellows which livened up what would
  otherwise have been a rather dull set of closing remarks. Finally, leadership
  requires funding and I thank again the European Community for funding the ELSA
  Marie Curie research training network through contract MRTN-CT-2006-033481.
\end{acknowledgements}


\end{document}